\documentstyle[11pt]{article}
\begin{document}

\def\sqr#1#2{{\vcenter{\hrule height.#2pt \hbox{\vrule width.#2pt
height#1pt \kern#1pt \vrule width.#2pt} \hrule height.#2pt}}}
\def\sq{{\mathchoice\sqr55\sqr55\sqr{2.1}3\sqr{1.5}3}\hskip 1.5pt}
\def\and{;\hfil\break}

\newcommand{\ECM}{\em Departament d'Estructura i Constituents de la
Mat\`eria\\
and I.F.A.E.
                  \\ Facultat de F\'\i sica, Universitat de Barcelona \\
                     Diagonal 647, 08028 Barcelona, Spain}

\def\thefootnote{\fnsymbol{footnote}}
\pagestyle{empty}
{\hfill \parbox{6cm}{\begin{center} UB-ECM-PF 96/7\\
                                    hep-ph/9602123 \\
                                    February 1996
                     \end{center}}}
\vspace{1.5cm}

\begin{center}
{\bf Wilsonian vs. 1PI renomalization group flow irreversibility}

\vskip .6truein
\centerline {Jordi COMELLAS\footnote{e-mail: comellas@sophia.ecm.ub.es} and
 Jos\'e Ignacio LATORRE\footnote{e-mail: latorre@sophia.ecm.ub.es}
}
\end{center}
\vspace{.3cm}
\begin{center}
\ECM
\end{center}
\vspace{1.5cm}

\centerline{\bf Abstract}
\medskip
We present a
line of reasoning based on the analysis of 
 scale variations of the Wilsonian partition function and  the
trace of the stress tensor in a curved manifold which results
in a 
statement of irreversibility of Wilsonian renormalization group flow for 
unitary  theories. We also analyze  subtleties  
related to subtractions in the case of
the 1PI effective action flow.

\newpage
\pagestyle{plain}

{\Large \bf Note}

An error in section 2 (which does not affect sections 3 and 4) invalidates
the flow equation for $c_W$. The paper has been temporarily withdrawn from
publication.

\section{A bit of history}
 \label{history}

Loss of information seems inevitable
as short-distance degrees of freedom are integrated
out in favor of an effective long-distance description of any physical
system. This intuition, which amounts to a statement of irreversibility of
renormalization group (RG) trajectories, might be deceptive. 
A renormalization group
transformation is made out of two steps: integration of modes (Kadanoff
transformation) followed by a rescaling of 
the variables of the system. Their combination
is not obviously irreversible, in particular when a  reshuffling of 
the Hilbert space comes along with the long distance realization of
the theory.

The efforts to  sort out this question started with
 the perturbative analysis of $\lambda \phi^4$ with $N$-components 
 made by Wallace and 
Zhia \cite{wallace}, proving that the beta functions of the model
 are indeed gradient flows
up to three loops.  The 
irreversibility conjecture was proven to be correct in that
context and, furthermore,  it was immediate to realize that 
\begin{equation}
\label{gradient}
\beta_i=\partial_i c\quad \longrightarrow \quad 
\int_{g^*_{UV}}^{g^*_{IR}} dg^i \
\beta_i\ =\Delta c \ge 0,
\end{equation}
where $c=c(g^i)$ is the function from which $\beta_i$ are derived and 
$g^*_{UV}, g^*_{IR}$ label 
the UV and IR fixed points delimiting the RG flow.
Thus, a sufficient, but not necessary, 
 condition for irreversibility of RG
flows is that the $\beta$-functions of a theory are gradient flows. What 
 is just necessary to prove is that an observable quantity existing in
any theory decreases monotonically along the RG flows.

With the advent of conformal field theories, Zamolodchikov constructed
a function that decreases along RG trajectories
 in two-dimensional unitary field theories
\cite{Zamolodchikov}. This function  reduces to
the central charge of the conformal theory at fixed points
(thus, Zamolodchikov's result is often refered as the c-theorem).
The elements of Zamolodchikov's proof were Lorentz invariance, conservation 
of the stress tensor and  unitarity. 

Later on, some groups have forcefully tried to extend Zamolodchikov's
 remarkable achievement to higher dimensional theories. Cardy
\cite{cardy} considered
the flow of the integrated trace of the stress tensor on a sphere, $S^n$,
as a candidate of a c-funtion.
 His idea consisted in trading the 
explicit subtraction
point dependence $(\mu)$  for the one in the  radius of the sphere $(a^{-1})$,
\begin{equation}
\label{cardyidea}
c(g(t),t)\equiv \langle \int_{S^n} \sqrt{g(x)} \theta(x) \rangle ,
\end{equation}
where $t\equiv \ln{\mu\over a}$, so that $\mu\partial_\mu=-a\partial_a$, 
where $a$ is the inverse of the radius of the sphere.
A RG transformation, usually unsderstood as an enlargement of all relative
distances, is here realized as a blow up of the sphere. 
 The RG flow of this 
c-candidate is  related to the correlator of two stress tensors, yet
the appearance of other contributions 
seemed to spoil the irreversibility proof. 
An appealing feature of Cardy's idea is that his candidate reduces
to the Euler density coefficient of the trace anomaly at conformal field 
theory \cite{duff}.  For instance, in four dimensions,
\begin{equation}
\label{traceanomaly}
\langle \theta(x)\rangle_{cft} = \frac{1}{2880}\left(
-3 a F(x) + b G(x) + c \,\sq R(x)\right)
\quad\Longrightarrow \quad\langle\int \sqrt{g}\theta\rangle_{cft}=
\frac{1}{2880}\,32\pi^2b\chi(M),
\end{equation}
where $F=C^2_{\mu\nu\rho\sigma}=R^2_{\mu\nu\rho\sigma}-
2R^2_{\mu\nu}+\frac{1}{3}R^2$,
$G=R^2_{\mu\nu\rho\sigma}-4R^2_{\mu\nu}+R^2$ and $\chi(M)$ the Euler
characteristic of the manifold.
Whereas $a$ has been proven to be related to the 
positive coefficient of 
the spin 2 structure in the correlator of two stress tensors in flat space
\cite{cfl}, 
$b$ is only empirically known to be positive for conformal free 
bosons, fermions and vectors
(its actual value is 1, 11 and 62, respectively). This $b$ coefficient
 as a candidate was further analyzed
and modified by Osborn  \cite{osborn} proving an evolution
equation reminiscent of the two-dimensional case. Unitarity did not
enter the construction and, again, irreversibility was not proven.

A different approach was taken in ref.~\cite{cfl,clv} where
the starting point was unitarity  via the spectral representation
of two stress tensor correlators. A reformulation of Zamolodchikov's
result
is achieved considering
\begin{equation} 
\label{spreptwodim}
\langle T_{\alpha\beta} (x) T_{\rho\sigma} (0) \rangle ={\pi\over 3}
   \int_0^\infty  d \nu \, c(\nu,t) \int {d^2 p\over (2\pi)^2}\,
      {\rm  e}^{i p x}{
	(  g_{\alpha\beta} p^2 -p_\alpha p_\beta) ( g_{\rho\sigma} p^2 - p_\rho
	p_\sigma) \over p^2 + \nu^2}\quad.
\end{equation} 
The spectral
function describes the central charge of UV and IR fixed points
associated to a given flow as
 \begin{equation} 
\label{cmu}
c(\nu,t)=c_{IR}\, \delta(\nu) +c_{smooth}(\nu,t) \qquad c_{UV}=\int
d\nu \ c(\nu,t)
 \end{equation}
where $t$ is the RG flow parameter.
 Therefore, $c_{UV}\ge c_{IR}$. At long
distances, only massless modes survive,
 entering the spectral representation as a delta of the spectral
parameter. Unitarity yields positivity of $c_{smooth}$, thus,
irreversibility. This proof emphasizes the role of the stress tensor as
a mean to account for all degrees of freedom, and to quantify the
decoupling of massive ones along RG trajectories. The extension of this
idea to higher dimensions encounters  a first complication due to the
existence of two spin
 structures in the $\langle TT\rangle$ correlator. Since RG flows are
related to a change of scale, it is natural to focus attention in the
spin 0 spectral density.  The problem turns out to be that, at
conformal field theories in $n$ dimensions, 
\begin{equation}
\label{conformal} c^{(0)}(\nu)\longrightarrow c^{(0)} \nu^{n-2}
\delta(\nu) 
\end{equation}
 and $c^{(0)}$ is unobservable, at least as a
well-defined quantity of the conformal field theory. The way out
proposed was to define $c^{(0)}$ using a limit from 
the flow itself 
\begin{equation}
 \label{wayout}
 c^{(0)}\delta(\nu) \equiv
\lim_{t\to 0} {c^{(0)}(\nu, t)\over \nu^{n-2}} \quad .
\end{equation}
 Any limiting procedure which
allows for a definition of $c^{(0)}$ yields a decreasing quantity. The
c-theorem is then proved in any perturbative expansion. These results
were verified through examples in ref.~\cite{clv}. 

Further efforts have been devoted to understand the validity of the
above proposed candidates
as well as to formulate the theorem by other means
\cite{hklm}. Let us here
summarize what we think are essential ingredients of the statement of
irreversibility of RG flows and, consequently, ought to enter its proof in one
way or another. We reduce such essentials to two:
\par {\sl a)} The c-function must be sensitive to all degrees of freedom
of the theory.
\par {\sl b)} The sign of the derivative of the c-function with respect to 
the flow parameter must be dictated by unitarity.

Therefore, it is not essential that the c-function reduces to
any simple quantity known in the theory, although it would be welcome
that such a possibility were realized. Neither much emphasis should be
put on the fact that beta-functions are gradient of the c-function.

Upon a simple reflection, it is clear that all candidates so far explored,
including Zamolodchikov's original one in two dimensions, are constructed
from the stress tensor. This operator exists for every theory and couples
to {\sl all} degrees of freedom. Furthermore, it does not pick anomalous
dimensions and 
its correlators obey useful Ward identities. It is not yet settled whether the
attemps to formulate the theorem in higher dimensions have so far 
missed some basic ingredient. On one hand,
the integrated 
trace of the energy momentum tensor, which carries  spin 0,
 generates a change  of scale, but
its variation has not been related to unitarity. 
On the other hand,
 the spectral representation approach
lacks the explicit formulation of a candidate at the conformal theory.
Yet the above requisits may be too stringent.
 The lesson from two
dimensions is that $c$ counts massless degrees of freedom whereas $\Delta c$
quantifies the decoupled massive modes. This modifies a little bit the
above point a). Similarly, we may also bring down unitarity requirements
of point b)
in the following way. If the c-function is stationary at fixed points,
it is just sufficient to find a negative sign of its derivative anywhere
along the flow. Then, the function will remain decreasing till a new
fixed point is reached.

We here propose to combine all this accumulated knowledge in the following way.
There is an obvious quantity in any field theory which is sensitive to all
degrees of freedom: the Wilsonian  partition function. We define it 
(in euclidean space)
 as
\begin{equation}
\label{partition}
Z[g^i(t),t]\equiv \int \prod_{\Lambda_{IR}\le p\le \Lambda_{UV}} d\varphi_p\,
e^{-S},
\end{equation}
where $t\equiv \ln{\Lambda_{UV}\over\Lambda_{IR}}$ is the RG flow parameter.
As more degrees of freedom are integrated out, $t\rightarrow \infty$. Often,
we consider field theories with the purpose of computing Green functions. Then,
the renormalization procedure trades $\Lambda_{UV}$ for a subtraction point
$\mu$ and $\Lambda_{IR}$ can be safely sent to 0 if external momenta
are different from zero (otherwise,  an IR regulator is 
necessary). We shall come back to this 
generating function for connected Green functions later on.

The absence of external sources shields the properties
of the partition function
from a simple analysis. We have found most convenient to consider field
theories defined on a curved space, {\sl e.g.} on the $S^n$ sphere with radius
$a^{-1}$. We may consider $a$ to be arbitrarily small, so that we use it
both as an external source for the stress tensor and as an IR regulator, taking
over the role of $\Lambda_{IR}$. Stress tensor correlators are defined as
 \cite{cfl},
\begin{equation}
\label{stresscorruno}
\langle \sqrt{g} T^{\mu\nu}(x)\rangle = - 2 V
{\partial \ln Z\over\partial g_{\mu\nu}(x)} ,
\end{equation}
\begin{equation}
\label{stresscorrdos}
\langle \sqrt{g}
T^{\mu\nu}(x) \sqrt{g}T^{\alpha\beta}(y)\rangle
 =  4 V^2 
{\partial \over\partial g_{\mu\nu}(x)}   
{\partial \over\partial g_{\alpha\beta}(y)}\ln Z + {V\over n} \langle 
\theta\rangle
 \delta^n(x-y)\sqrt{g}
\left(g^{\mu\nu} g^{\alpha\beta}- 
g^{\nu(\alpha} g^{\mu\beta)}\right) ,
\end{equation}
where $V={2 \pi^{n+1\over 2}\over \Gamma({n+1\over 2})}$ is the 
standard volume
factor of a sphere.
A simple definition of these correlators in terms of functional
derivatives with respect to the metric would lead to a violation
of diffeomorphism Ward Identities at contact terms. This is the reason
to subtract a delta term in eq.~(\ref{stresscorrdos}), which has
been further simplified using properties of maximally symmetric spaces.
As it stands, both $\nabla_\mu\langle T^{\mu\nu}\rangle=0$ and
$\nabla_\mu\langle T^{\mu\nu}T^{\alpha\beta}\rangle=0$ are obeyed.
Further properties of $T^{\mu \nu}$ are discussed
in ref.~\cite{tmunu,coste}.

We are now in the position of streamlining our construction.

\section{Irreversibility of the Wilsonian RG}

Consider a field theory defined on a $S^n$ sphere of radius $a^{-1}$. We
construct the following dimensionless quantity based on the Wilsonian
partition function
\begin{equation}
\label{candidate}
c_W(g^i(t), t) = \int \sqrt{g(x)} \langle \theta(x)\rangle + n V 
\ln Z ,
\end{equation}
where $t=\ln{\Lambda\over a}$, $\Lambda$ being an UV scale.
This function obeys the RG equation
\begin{equation} 
\label{RGequation}
\left( {\partial\over \partial t} + \beta_i {\partial\over \partial g^i}
\right) c_W(g^i(t),t)=0.
\end{equation}
We need to study the RG flow of $c_W$, that is
\begin{equation}
\label{RGflow}
{\partial\over \partial t} c_W= -\beta_i {\partial\over \partial g^i}
 c_W .
\end{equation}
The $t$ dependence can be computed as 
\begin{equation}
\label{tradeoftfora}
{\partial\over \partial t}=-a {\partial\over \partial a} ,
\end{equation}
which reflects the fact that changing the scale at which physics is
considered can be done by varying the radius of the sphere. This is
nothing else than the starting point of Cardy's analysis.
In a symmetric space it is also true that
\begin{equation}
\label{tradeofaforg}
a {\partial\over \partial a}=-2\int g_{\mu\nu}(x)
{\delta\over \delta g_{\mu\nu}(x)},
\end{equation}
a change in the scale factor is obtained as a Weyl transformation.

Using the definitions in eq.~(\ref{stresscorruno}, \ref{stresscorrdos}) 
we, first,  have  that
\begin{equation}
\label{trace}
\sqrt{g(x)}\langle\theta(x)\rangle=-2 V g_{\mu\nu}(x)
{\delta\over \delta g_{\mu\nu}(x)}\ln Z,
\end{equation}
and, second,
\begin{equation}
\label{tracetrace}
\sqrt{g(x)}\langle\theta(x)\theta(y)\rangle=-2 V g_{\mu\nu}(x)
{\delta\over \delta g_{\mu\nu}(x)}\langle\theta(y)\rangle.
\end{equation}
As a particular case, these relations control the trace anomaly at conformal
field theory. The trace of the energy momentum tensor does not vanish,
neither its correlators at coincident points, although $\beta_i(g^j{}^*)=0$.
Putting together all the ingredients
\begin{equation}
\label{summary}
{\partial\over \partial t}c_W=
-a {\partial\over \partial a}c_W=2\int g_{\mu\nu}(x)
{\delta\over \delta g_{\mu\nu}(x)}c=
-{1\over V} \int_x\int_y 
\sqrt{g(x)}\sqrt{g(y)}\langle\theta(x)\theta(y)\rangle
\end{equation}
Thus,
\begin{equation}
\label{cflow}
{\partial\over \partial t}c_W= -{1\over V} \langle\int_x 
\sqrt{g}\theta(x)\int_y\sqrt{g}\theta(y)\rangle\le 0 .
\end{equation}
The c-function we have constructed from the Wilsonian partition function
decreases along RG flows. All intermediate steps are well-defined as 
the partition function is equipped with both an UV and IR cut-off at 
the outset. This obviates problems related to subtractions or non-analiticity
at zero momentum, in close analogy to the currently widely used idea
that the Wilsonian effective action is analytic in momenta. The exact
renormalization group equation can be expanded in momenta due to the
ubiquous presence of cut-offs. It is a delicate issue how to handle the
case of the flow for the 1PI effective action, which we postpone to the
next section.

Our candidate in eq.~(\ref{candidate}) provides a synthesis of several
ideas previously considered. It takes Cardy's conjecture as a starting point,
bringing a piece of the needed irreversibility together with a term which
spoils it. This is corrected with the introduction of the log of the partition
function. Some authors  had conjectured that the 
partition function itself would be enough to provide a c-function. Such
a possibility is definitely ruled out in flat space as easily seen in the
example of a free boson plus a free fermion. The overall partition function
cancels, therefore the piece coming from  a free boson flows decreases
as the one for a free fermion increases. It also follows that a naive
use of the exact renormalization group equation to settle irreversibility
of the flow is insufficient. The equation controls the flow of the
partition function  which does not decrease monotonically as we just
pointed out.

\section{1PI effective action flow}

We would like to explore the limit of the flow equation we have gotten in
the previous section to the case  where $\Lambda_{UV}$ is sent to infinity.
This is carried through a standard renormalization procedure that 
trades $\Lambda_{UV}$ for a subtraction point $\mu$, often 
sending also $\Lambda_{IR}\to 0$, when external momenta of Green functions
are kept different from zero. This limit of the partition function leads,
upon a legendre transformation, to 
 the 1PI effective action. For convenience, we  shall call it 1PI partition
function.
 The Wilsonian partition function flow 
interpolates between two theories, but loops associated to the second, IR,
theory remain to be done. In the 1PI effective action we do integrate all
modes.

From the point of view of our irreversibility argument,  two issues
need to be reconsidered. The first one, $\Lambda_{IR}\to 0$
 is bypassed since the curvature
provides a natural IR cut-off. To keep $b$ as an essential ingredient of
 the c-function at conformal points, it is necessary to work in curved space.
 Therefore, we  concentrate on the UV limit.

Let us first note that
the stress tensors correlators are defined from the renormalized partition 
function. Standard coupling constants and wave-function renormalizations have
been performed. Furthermore, the stress tensor 
combines the wave-function renormalization of its constituent fields with a 
composite operator one such that $\theta_{bare}=
\theta_{ren}$
and, thus, 
carries no anomalous dimensions.
What remains is just a subtraction
in the two-point  correlator
 $\langle T_{\mu\nu}(x) T_{\alpha\beta}(0)\rangle$.
 It is convenient to understand 
this point in terms of
the freedom of scheme brought by the renormalization procedure.
In four dimensions,
the above correlator has dimension 8, but Ward Identities dictate the
presence of 4 derivatives which leaves a freedom of a contact term.

On a sphere of radius $a^{-1}$, the spectral representation takes
the form
\begin{equation}
\label{specrepsn}
\langle \theta(x)\theta(0)\rangle=
\frac{\pi^2}{40}a^2\int_{\frac{5}{2}}^{\infty}d\sigma
\,\rho(\sigma)\left(\Delta-4 a^2\right)^2G(\sigma,r),
\end{equation}
where $\sigma$ labels the scalar representations of $SO(1,4)$,
$\Delta$ is the covariant
laplacian, $r$ the geodesic distance and $G(\sigma,r)$ the
apropriate Green
function, $\left[\Delta-a^2(\sigma^2-\frac{9}{4})\right]G(\sigma,r)
=\frac{\delta^4(x)}{\sqrt{g(x)}}$.
At CFT, $\rho(\sigma)=\rho_0\delta(\sigma-\frac{5}{2})$.
A detailed analysis
\cite{cfl} reveals the presence of a contact term
\begin{equation}
\label{contactterm}
\langle \theta(x)\theta(y)\rangle =
\frac{\pi^2}{40}a^2\rho_c \left(\Delta-4 a^2\right)
\frac{\delta^4(x-y)}{\sqrt{g(x)}}.
\end{equation}
Using eq.~\ref{tracetrace}, one proves that
this contact term is related to the coefficient of the spectral function 
at the conformal point,
$\rho_0\ge  0$,
and the $b$ trace anomaly coefficient in the
following way \cite{cfl}
\begin{equation}
\label{relation}
b=\rho_0+\rho_c.
\end{equation}
Through examples, $\rho_c$ at most cancels $\rho_0$ (in odd dimensions)
but never overcomes it, leaving $b\ge 0$ always. We have a weak argument
for this result. Take a resolution of the identity on the r.h.s.~of 
(\ref{cflow}),
\begin{equation}
\label{resolution}
- \sum_n\langle \int\sqrt{g}\theta \vert n \rangle 
\langle n\vert \int\sqrt{g}\theta\rangle \le 0 .
\end{equation}
A unitarity argument of this sort, if applicable on integrated objects,
yields irreversibility for the 1PI effective action. The above argument
is protected from IR infinities in $S^4$.
This reasoning would preserve the sign of the r.h.s. in eq.~(\ref{cflow})
in the 1PI case at the same time that would 
also explain the elusive positivity of $b$.

\section{Review of  Cardy's proposal}

As discussed earlier, a theorem stating irreversibility of RG
trajectories
 just needs proving a definite sign for the derivative of 
an observable quantity on the flow parameter. This, though, stays
one step short from a powerful quantitative tool if  there is no simple 
way to characterize the c-function. The  discussion 
in the two previous sections is missing
such a point. We already noted  that
 the welcome properties of a c-function
are those offered by the original two-dimensional case:
the c-function should be stationary at conformal field theory;
at these points, the c-number should be easily computable;
and, in the best of the worlds,
beta-functions should be gradient flows.

The understanding of contact terms in stress tensor correlators, as
explored in the previous section,  
allows for a more detailed analysis of a simpler candidate for the c-theorem. 
Let us reconsider Cardy's proposal of eq.~(\ref{cardyidea})
in dimension 4. Its variation
along the flow can be derived using the equations in section 2,
\begin{equation}
\label{cardyflow}
{\partial\over\partial t} c(g^i(t),t)=-{1\over V} \int_x
\int_y \sqrt{g(x)}\sqrt{g(y)} \langle \theta(x)\theta(y)\rangle
+n \ c(g^i(t),t).
\end{equation}
This quantity is not obviously negative due to the presence of the last term.
One can trace the two terms in the r.h.s. to the variation of $\theta$
and $\sqrt{g}$ respectively. In a way, the competition of these two terms
is forced by the adimensional character of $c$. 
At conformal field theory, this c-function reduces to the $b$ coefficient
of the trace anomaly and is stationary,
\begin{equation}
\label{stationarity}
\left.
{\partial\over \partial t} c\right|_{g^i=g^i_*}={\partial\over \partial t}
b=0,
\end{equation}
because the r.h.s.~yields and exact cancellation between the contact
term coming from the two-point correlator and the trace anomaly,
as derived from eq.~(\ref{tracetrace}).

 It is clear that $\partial_t c=0$ at fixed points implies that
it is sufficient to find a negative sign of $\partial_t c$ at any intermediate
point to have a proof of irreversibility. This, indeed, was done
by Cardy near a fixed point using conformal perturbation theory
for quasi-marginal deformations.
His computation builds more confidence of the validity of the theorem but
is still not a proof. 

Let us go back to the above observation about the exact cancellation
of the r.h.s. of eq.~(\ref{cardyflow}) at fixed points.
 The origin of this simplification 
is rooted in the operator product expansion of $\theta(x)\theta(0)$. 
It is known that the identity contribution in the flat space OPE is
\cite{duff,cfl,coste}
\begin{equation}
\label{flatope}
T_{\mu\nu}(x) T_{\alpha\beta}(0)\sim c^{(2)} \Pi^{(2)}_{\mu\nu,\alpha\beta} 
(\partial){1\over x^4} + \dots 
\end{equation}
where $\Pi^{(2)}_{\mu\nu,\alpha\beta}(\partial)$
 stands for a spin 2 projector and $c^{(2)}$ 
can be shown to coincide with the $a$ coefficient of 
the Weyl square density in the trace anomaly (\ref{traceanomaly}).
This, indeed, provides a
connection between spin 2 flat space non-local correlators and 
spin 0 contact terms in curved space.
In curved space, the OPE for the
traces of stress tensors contains a delta term
\begin{equation}
\label{OPE}
\theta(x)\theta(0)\sim {4\over V} \theta(0) \delta^4(x) +
\mbox{\rm non-local terms} 
\end{equation}
Indeed, this contact term is needed in the OPE for consistency of the
anomaly in curved space. The factor $4$ is related to the classical (as
well as quantum) dimension of $\theta$, whereas the factor $V$ is present
due to our definition of $\theta$. No other contact terms are present 
by dimensional and scaling arguments. The global prefactor is 
fixed by the way the OPE works in the conformal case. Away from
the fixed point, we have so far no control on the relation between the
two terms in the r.h.s.~of eq.(\ref{cardyflow}). 

The above analysis adds some understanding to the RG flow of the Euler density
coefficient in the trace anomaly but does not prove its irreversibility yet.
More inconclusive but tantalizing evidence was presented through an 
example by Cardy.
Consider QCD at short and long distances. Asymptotic freedom allows
for an easy computation of $c_{UV}= 11 N_f N_c + 62 (N_c^2-1)$ whereas
its chiral realization implies $c_{IR}= (N_f^2-1)$, since
$b=1,11,62$ for bosons, fermions and vectors. It follows that
$c_{UV}>c_{IR}$. Moreover, in ref.~\cite{bastianelli}, a large number
of exact results have been checked against the above c-function
and  systematic validity of the would-be theorem has been found.

One deep, uncanny lesson hidden in this ideas is that fermions
{\sl weight} more than bosons. This is not so in two dimensions
as a Dirac fermion has the same central charge as a boson, which
is at the origin of exact bosonization. This is no longer true in higher
dimensions. Long distance realizations would, in general, favor bosons.
This might represent just a glimpse of a deep reletation 
between RG irreversibility
and Goldstone theorem.

\section{Acknowledgments}

We are indebted for the  insights shared in countless 
discussions with
A. Cappelli, P.E. Haagensen, E. Moreno and P. Pascual.

Financial support from CICYT,
contract AEN95-0590, and CIRIT, contract GRQ93-1047, are also
acknowledged.

\end{document}